**PROSPECT-PRO for estimating content of nitrogen-containing leaf proteins and other carbon-based constituents**

Jean-Baptiste Féret[1], Katja Berger[2], Florian de Boissieu[1], Zbyněk Malenovský[3]

[1] TETIS, INRAE, AgroParisTech, CIRAD, CNRS, Université Montpellier, Montpellier, France

[2] Department of Geography, Ludwig-Maximilians Universität München, Luisenstr. 37, 80333 Munich, Germany

[3] Department of Geography and Spatial Sciences, School of Technology, Environments and Design, College of Sciences and Engineering, University of Tasmania, Private Bag 76, Hobart 7001, Australia

**_Highlights_**

- Leaf dry matter is decomposed into proteins and carbon-based constituents in PROSPECT-PRO

- The specific absorption coefficient of proteins revealed expected absorption features

- Inversion of PROSPECT-PRO accurately estimated foliar protein content of dry and fresh leaves

- Leaf nitrogen content may be quantified by estimating leaf protein content with PROSPECT-PRO

- New PROSPECT-PRO model is fully compatible with its predecessor PROSPECT-D model

***Abstract***

Models of radiative transfer (RT) are important tools for remote sensing of vegetation, as they facilitate forward simulations of remotely sensed data as well as inverse estimation of biophysical and biochemical properties from vegetation optical properties. The remote sensing estimation of foliar protein content is a key to monitoring the nitrogen cycle in terrestrial ecosystems in particular to better understand photosynthetic capacity of plants and improve nitrogen management in agriculture. However, no physically based leaf RT model currently allows for proper decomposition of leaf dry matter into nitrogen-based proteins and carbon-based constituents (CBC), estimated from optical properties of fresh or dry foliage.

We developed a new version of the PROSPECT model, named PROSPECT-PRO, which separates nitrogen-based constituents (proteins) from CBC (including cellulose, lignin, hemicellulose and starch). PROSPECT-PRO was calibrated and validated on subsets of the LOPEX dataset, accounting for both fresh and dry broadleaf and grass samples. We applied an iterative model inversion optimization algorithm to identify optimal spectral subdomains for retrieval of leaf protein and CBC contents, with 2125-2174 nm optimal for proteins and 2025-2349 nm optimal for CBCs. PROSPECT-PRO inversions revealed a better performance in estimating proteins from optical properties of fresh than dry leaves (respectively, validation $R^2$=0.75 and 0.62, NRMSE= 17.3% and 24%), but similar performances for the estimations of CBCs (respectively validation $R^2$=0.92 and 0.95, NRMSE= 13.01% and 13.9%).

We further tested the ability of PROSPECT-PRO to estimate leaf mass per area (LMA) as the sum of proteins and CBC using independent datasets acquired for numerous plant species. Results showed that PROSPECT-PRO is fully compatible and comparable with its predecessor PROSPECT-D in indirect estimation of LMA, with validation $R^2$=0.90 and NRMSE=16.3% for PROSPECT-PRO and $R^2$=0.91 and NRMSE=18.6% for PROSPECT-D across eight independent data sets. We can conclude from findings of this study that PROSPECT-PRO has a high potential in establishing the carbon-to-nitrogen ratio based on the retrieved CBC-to-proteins ratio ($R^2$=0.89 and NRMSE=12.7% for fresh leaves and $R^2$=0.58 and NRMSE=30.7% for dry leaves). That might be of interest for precision agriculture applications estimating carbon and nitrogen from observations of current and forthcoming airborne and satellite imaging spectroscopy sensors.

1. INTRODUCTION

Nitrogen (N) is a major nutrient for all living plant organisms, cultivated as well as wild forms. In agriculture, crop yield quality is primarily dependent on protein content, with the N availability being the most critical factor of actual grain protein content (Brown et al., 2005). N limitation in soil and plants generally restricts development and growth of roots, suppresses lateral root initiation, increases the carbon-to-nitrogen (C:N) ratio within the plant, reduces photosynthesis, and results in an early leaf senescence (Kant et al., 2011; Paul and Driscoll, 1997; Wingler et al., 2006). On the other hand, N over-fertilization is undesirable for quality of both crops and environment. Excess of N reduces yield and decreases its quality (e.g. organoleptic quality), reduces content of mineral nutrients and secondary metabolites, and increases nitrate content in leaves (Albornoz, 2016). From an environmental perspective, human activity altered the global N cycle by fertilization of farming systems increasing the fixation rate of N on land, which has negative consequences on terrestrial and aquatic ecosystems (Davidson et al., 2011; Gruber and Galloway, 2008). The consequences include habitat eutrophication, acidification, and contribution to the accelerated loss of biodiversity caused by decreased competitive advantage of plants adapted to efficient use of nitrogen (Vitousek et al., 1997). Optimization of N management is, therefore, of high importance to mitigate these negative consequences, while securing sufficient and sustainable food production. More generally, N concentration in all plants, both agriculture and not, is generally considered as an important surrogate measure for plant photosynthetic capacity (Evans, 1989), and its remote estimation is, therefore, of a great interest more broadly in plant biology and ecology.

Remotely sensed (RS) monitoring of N in vegetation is a prospective tool for improvement of N management and reduction of negative impacts imposed by conventional farming. Decision support systems integrating RS information are usually based on the indirect empirical relationship between leaf N and chlorophyll content. Such monitoring has some operational advantages, originating from strong chlorophyll a+b spectral absorption features in the visible domain and a great diversity of physically based, data driven and hybrid methods designed to estimate chlorophylls from multi- and hyperspectral data (Baret et al., 2007; Clevers and Gitelson, 2013; Malenovský et al., 2013; Verrelst et al., 2015). Although a significant amount of literature reported a strong correlation between leaf N and chlorophyll content in crops (Baret et al., 2007; Clevers and Kooistra, 2012; Vos and Bom, 1993; Yoder and Pettigrew-Crosby, 1995), this relationship does not hold during their senescence and does not appear to be universal, as it is relatively weak across species and ecosystems (Asner and Martin, 2009; Homolová et al., 2013). N is involved in many leaf physiological processes, including photosynthesis, respiration, structural growth and storage capacity building (Liu et al., 2018). This results in multiple N based leaf biochemical constituents in different physiological roles throughout the plant cycle and in response to environmental factors. Chlorophyll pigments contain only a small fraction of N in plants, representing less than 2% of the total leaf N (Kokaly et al., 2009). In comparison, proteins are the major nitrogen-containing biochemical constituents, with Ribulose-1,5-bisphosphate carboxylase/oxygenase (rubisco enzyme) holding 30–50% of the N in green leaves (Elvidge, 1990; Kokaly et al., 2009). Rubisco, the most abundant protein on Earth, catalyzes the photosynthetic fixation of carbon dioxide (Sharwood, 2017) and together with other photosynthesis-related proteins the major source of N available for remobilization among plant parts (Masclaux-Daubresse et al., 2010). For instance, N in oilseed rape (*Brassica napus*) remobilizes from senescing to expanding leaves during the vegetative growth stage and from senescing leaves to seeds during the reproductive stage (Malagoli, 2005). This indicates that, unlike chlorophyll, plant nitrogen content does not decrease upon reaching mature growth stages, but is rather translocated to other organs, which makes the relationship between plant nitrogen and leaf chlorophyll content nonlinear through the vegetation growth cycle. Consequently, a quantitative non-destructive retrieval of leaf protein content should be a more reliable proxy of nitrogen content (Berger et al., 2020).

As reported in the pioneering studies from Curran (1989), Elvidge (1990) and Himmelsbach et al. (1988), the absorption features corresponding to proteins are caused mainly by N-H bond stretches. These absorption features are located in the shortwave infrared (SWIR) domain between 1500 and 2400 nm, with two additional absorption features reported in the near infrared (NIR) domain at 910 and 1020 nm. The quantification of proteins from leaf optical properties (LOP) or a canopy reflectance factor is, however, challenging, because of their relatively low concentrations, and because their specific absorption features are masked by water absorption and overlapped by absorption features from other dry matter constituents (Fourty et al., 1996; Jacquemoud et al., 1996). Multispectral optical systems with broad spectral bands and moderate spectral sampling are insufficient to correctly differentiate biochemical constituents with narrow and overlapping absorption features. The contiguous narrow spectral bands measured with leaf spectroscopy and imaging spectroscopy are more suitable to differentiate spectral features corresponding to the combination of multiple optically active constituents (Hank et al., 2019). Even the subtle contribution of proteins to the hyperspectral signal may allow their accurate estimation, if using appropriate methods. Methods for such task include multivariate statistical and machine learning algorithms, physically based approaches or hybrid combinations of both (Verrelst et al., 2019a).

Physical models offer a certain number of advantages over empirical and machine learning approaches. The physically explicit representation of the interactions between radiance and vegetation enables forward simulation and inversion of reflectance signals acquired by any, even future, laboratory/field, close-range, airborne or space-borne spectroradiometer. The advantages in terms of robustness and transferability are substantial compared to data-driven methods, although recent publications suggest that differences may not be as large as expected (Serbin et al., 2019). Alternatively, definition of a physical model that includes proteins as an input requires the calibration of specific absorption coefficients (SACs) for proteins and other constituents as dry matter in leaves, which has proven to be challenging (Botha et al., 2006; Kokaly et al., 2009). Jacquemoud et al. (1996) developed a version of the PROSPECT model including SAC for proteins and for different combinations of carbon-based constituents (CBCs), but the model inversion resulted in moderate to good estimates of proteins ($R^2$ between 0.49 and 0.67) and different combinations of CBC ($R^2$ between 0.39 and 0.88) when applied to dry leaves and poor to moderate accuracy for proteins ($R^2<0.05$) and CBCs ($R^2<0.50$) when applied to fresh leaves. Wang et al. (2015) also developed an updated version of PROSPECT including proteins and lignin + cellulose. Following the calibration procedure of Féret et al. (2008), they concluded on the importance of selecting specific spectral domains to obtain optimal results, which was later confirmed by Féret et al. (2019) when inverting PROSPECT for estimation of leaf dry mass per area (LMA) and equivalent water thickness (EWT). However, both Jacquemoud et al. (1996) and Wang et al. (2015) assumed that only proteins, lignin and cellulose, representing about 75% of LMA, contribute to leaf absorption. They excluded spectral contribution of non-structural carbohydrates (e.g. glucose and starch), which may result in a significant source of forward and inverse modelling uncertainties.

Our overall objective is to develop a new version of the PROSPECT model capable of differentiating and accurately estimating protein and CBC contents from leaf spectroscopic measurements. The model should be fully compatible with its previous version PROSPECT-D, i.e. able to simulate the spectral contributions of proteins and CBCs consistently with the original single LMA input, including an accurate LMA retrieval through the model inversion. The new PROSPECT version, named PROSPECT-PRO, should be applicable to all types of bifacial leaves, including fresh green as well as senescent and dry leaves. As a secondary objective, we intend to identify optimal spectral domains for quantitative estimation of leaf proteins through a PROSPECT-PRO inversion and validate its performance on independent datasets of leaf optical and biochemical measurements. The introduced improvements in PROSPECT-PRO leverage only shortwave infrared (SWIR, 1000-2500 nm) wavelengths, where protein and CBC absorption features are prominent, and therefore does not affect existing functionality of PROSPECT with respect to foliar pigments.

We provide a general introduction to the PROSPECT model physical principles in Section 2. The data used for the calibration and validation of PROSPECT-PRO are described in Section 3, followed by the calibration procedure, including analytical tools for global sensitivity analysis, validation and identification of optimal retrieval spectral domains, described in Section 4. Section 5 presents the results of the study. Finally, Section 6 discusses potential applications of and limitations to PROSPECT-PRO, with concluding findings in Section 7.

2. General presentation of PROSPECT

PROSPECT is a physical model simulating leaf directional-hemispherical reflectance and transmittance (Schaepman-Strub et al., 2006) using a relatively low number of biophysical and biochemical input parameters. Its first version was developed by Jacquemoud and Baret (1990) as an extension of the generalized plate model of Allen et al. (1970, 1969), with later versions developed to include more absorbing constituents (Jacquemoud et al., 1996; Féret et al., 2008, 2017) or to adapt to specific conditions and leaf types, for example needle-shaped leaves (Malenovský et al., 2006). The PROSPECT model was also the starting point for development of independent extensions modelling RT of leaf chlorophyll fluorescence, such as FluorMODleaf (Pedrós et al., 2010) and Fluspect (van der Tol et al., 2019; Vilfan et al., 2018). PROSPECT can be used in forward mode to simulate leaf optical properties from the description of its biochemical and structural properties, or in inverse mode to estimate part or all of these biochemical and structural properties based on measured leaf optical properties. Detailed description of these modes is provided in Section 4.

In addition to the leaf biochemical variables such as pigments, EWT and LMA, PROSPECT requires a unique leaf mesophyll structure parameter $N_{struct}$. $N_{struct}$ corresponds to the number of uniform compact plates separated by $N_{struct} - 1$ air spaces within the leaf. The value of $N_{struct}$ represents the complexity of the leaf internal structure, where a low $N_{struct}$ corresponds to moderate mesophyll tissue complexity (e.g. most of Monocots) while a higher $N_{struct}$ indicates greater complexity (e.g. many Eudicots) (Boren et al., 2019). $N_{struct}$ governs leaf internal scattering, but it has only a minor impact on leaf absorption. Higher values of $N_{struct}$ increase internal scattering resulting in greater reflectance and decreased transmittance, which is expressed primarily in spectral domains of low absorption (e.g. the NIR). To date, no standard experimental protocol to measure $N_{struct}$ exists, except an indirect estimation from NIR leaf reflectance and transmittance measurements. Since we used the most recent model version PROSPECT-D as a basis for establishing a new PROSPECT-PRO, the wavelength dependent refractive index of leaf interior and the SACs corresponding to water remained identical to PROSPECT-D.

3. MATERIAL

a. Calibration and validation data to establish PROSPECT-PRO

First, we needed to identify a suitable calibration dataset including directional-hemispherical leaf reflectance and transmittance and corresponding biochemical destructive measurements of the constituents defined as model inputs. Only constituents with optical activity within the spectral range in which calibration is performed are needed, so content of foliar pigments were not required for calibration as the new additions in PROSPECT-PRO utilize the SWIR domain covering protein absorption features. The Leaf Optical Properties Experiment (LOPEX) dataset, established by the Joint Research Center (JRC) of the European Commission (Ispra, Italy) in 1993 (Hosgood et al., 1994), contains optical, physical and biochemical measurements of more than 50 plant species collected around Ispra, Italy. The optical properties of leaf directional-hemispherical reflectance and transmittance were measured with an integrating sphere from the visible (VIS) to shortwave infrared region (VSWIR, 400-2500 nm). The biochemical measurements of photosynthetic pigments, water (EWT) and generic dry matter (LMA) content, as well as carbon (C), hydrogen, oxygen, nitrogen, lignin, proteins, cellulose and starch content are expressed as a percentage of dry mass. The protein content in the original LOPEX dataset was estimated from the nitrogen content measured by the Kjeldahl method (Bradstreet, 1954; Sáez-Plaza et al., 2013) using the widely used nitrogen-to-crude protein conversion factor of 6.25, which is widely used for food materials. We used the revised factor of 4.43, as suggested by Yeoh and Wee (1994) to be more representative of a broader range of vegetation types.

As the original version of LOPEX included 120 samples, encompassing broad leaves, needles, stalks, and powders, we used data corresponding to bifacial monocotyledons and eudicotyledons. Five reflectance and transmittance measurements taken for each sample were averaged. For some samples, optical properties were measured from both fresh and dry leaves; we separated these to produce two distinct datasets of dry and fresh samples. Samples in the original LOPEX dataset that were collected to test repeatability were averaged. Chemical measurements were performed by two independent laboratories in Belgium and France (Verdebout et al., 1995). Although measurements of both laboratories were relatively consistent, we decided to use the chemical analyses from the Belgian laboratory, leading to slightly improved overall results during calibration and validation stages. One sample was discarded from the fresh samples, because its presence systematically prevented a proper calibration and validation across all tests of the data, and two fresh samples were placed in the validation data because they resulted in poor calibration. These three samples were all characterized by high EWT > 0.030 cm (i.e. 30 mg.cm$^{-2}$). Therefore, the final selection of the LOPEX dataset resulted in 66 fresh and 49 dry eligible samples.

To our best knowledge, LOPEX is the only open dataset that includes required information on leaf protein content for the calibration and validation of PROSPECT-PRO. Therefore, we split LOPEX into independent calibration and validation subsets. To minimize risks of an imbalanced distribution of protein content between calibration and validation sets, all dry and fresh samples were pooled together and rank ordered. Every second sample was selected for calibration and the remaining samples were used for validation. The calibration datasets will be referred to as CALIBRATION while the validation datasets will be identified as VALIDATION, and the combined dataset will be referred to as LOPEX-CALVAL. Dataset statistics are provided in Table 1, with correlations among chemical constituents of fresh samples presented in Section 5.a.

Table 1. Statistical summary of fresh and dry samples included in the CALIBRATION and VALIDATION datasets: mean values and range.

| Name | Samples | LMA (mg.cm$^{-2}$) | Proteins (mg.cm$^{-2}$) | Protein concentration (%DW) |
|---|---|---|---|---|
| CALIBRATION | | | | |
| Dry | 23 | 5.84 (2.35-9.07) | 0.78 (0.38-1.35) | 14.32 (7.31-25.22) |
| Fresh | 33 | 5.29 (2.58-13.69) | 0.66 (0.17-1.23) | 13.66 (5.02-26.06) |
| VALIDATION | | | | |
| Dry | 26 | 5.89 (2.55-16.58) | 0.77 (0.15-1.37) | 13.98 (5.02-26.06) |
| Fresh | 33 | 5.18 (1.88-10.88) | 0.69 (0.29-1.22) | 14.64 (6.97-28.94) |

b.    Data for compatibility assessment between PROSPECT-D and PROSPECT-PRO

We assembled a second dataset to test the capability of PROSPECT-PRO to estimate LMA as the combination of leaf proteins plus CBC contents in comparison to the previous version PROSPECT-D (Féret et al., 2017). For this, we included six additional datasets: ANGERS, HYYTIALA, ITATINGA, NOURAGUES, PARACOU and LOPEX-Full (Féret et al., 2019) (see Table 2 for EWT and LMA statistics). Note that LOPEX-Full includes all individual measurements of leaf optical properties corresponding to 330 measurements of LMA and EWT for fresh leaves (66 fresh samples and five repetitions for each sample, each of these repetitions corresponding to an individual set of leaf optical properties, LMA and EWT), while the LOPEX CALIBRATION and VALIDATION datasets contain averages of these five repetition, including their corresponding protein contents measured from the five leaf samples combined together (see Hosgood et al., 1994 for additional details). None of these six datasets include destructive measurements of leaf protein content.

Table 2. Statistical summary of additional experimental datasets used to test the compatibility between PROSPECT-PRO and PROSPECT-D: mean values and ranges of minimum – maximum measurements.

| Name | Samples | EWT (mg.cm$^{-2}$) | LMA (mg.cm$^{-2}$) |
|------|---------|--------------------|--------------------|
| ANGERS | 308 | 11.47 (4.40 − 34.00) | 5.12 (1.66 − 33.1) |
| HYYTIALA | 96 | 9.16 (3.68 − 23.73) | 6.27 (2.76 − 15.77) |
| ITATINGA | 415 | 14.44 (2.20 − 20.20) | 10.24 (6.90 − 14.70) |
| LOPEX-Full | 330 | 11.13 (0.29 − 52.49) | 5.30 (1.71 − 15.73) |
| NOURAGUES | 262 | 11.73 (3.20 − 38.10) | 10.81 (3.10 − 21.10) |
| PARACOU | 272 | *N/A (N/A − N/A)* | 12.32 (5.28 − 25.56) |

4. METHODS

    a. PROSPECT forward modelling and inversion

In forward mode, PROSPECT simulates LOP based on a set of biophysical and biochemical properties (leaf chemistry and $N_{struct}$). In inverse mode, the optimal set of biophysical and biochemical properties can be identified via a variety of methods, for example using a merit function that minimizes the difference between measured and simulated LOP. A common inversion procedure is based on the numerical minimization of the sum of weighted square errors over all available spectral bands (Baret and Buis, 2008;

Féret et al., 2019). The minimized merit function $M$, using both reflectance and transmittance properties, is expressed as follows:

$$M\left(N_{struct}, \{C_i\}_{i=1:p}\right) = \sum_{\lambda=\lambda_1}^{\lambda_n} \left[W_{R,\lambda} \times \left(R_\lambda - \hat{R}_\lambda\right)^2 + W_{T,\lambda} \times \left(T_\lambda - \hat{T}_\lambda\right)^2\right],$$

Eq. 1

where $N_{struct}$ is the leaf structure parameter, $p$ is the number of chemical constituents accounted for by PROSPECT and retrieved during the inversion, $C_i$ the biochemical content per unit of leaf surface for a constituent $i$, $\lambda_1$ and $\lambda_n$ are the first and last wavebands entering the inversion, $R_\lambda$ and $T_\lambda$ are the experimental reflectance and transmittance measured at waveband $\lambda$, $\hat{R}_\lambda$ and $\hat{T}_\lambda$ are the reflectance and transmittance simulated by PROSPECT with $\{N_{struct}, \{C_i\}_{i=1:p}\}$ as input variables, and $W_{R,\lambda}$ and $W_{T,\lambda}$ are the weights applied to the squared difference between experimental and simulated reflectance and transmittance, respectively. Eq. 1 can be used to estimate all input variables, or just their limited subset, if prior information or arbitrary values for some variables are known. Here, the values of $W_{R,\lambda}$ and $W_{T,\lambda}$ were set to 1, giving all wavelengths the same importance.

b. Calibration of PROSPECT-PRO

Our primary objective is to include proteins and CBC as new leaf constituents in PROSPECT-PRO, while maintaining compatibility with PROSPECT-D in both the accuracy of the forward simulated LOP and estimations of EWT and LMA. The spectral contributions of proteins to LOP in previous versions of PROSPECT were simulated through LMA (Féret et al., 2017, 2008), which contains also CBCs such as lignin, cellulose and non-structural carbohydrates (sugars and starches). Each of these CBCs is characterized by a specific carbon content (Ma et al., 2018), but contains no N. However, the distinction of each single constituent of LMA is beyond the scope of this study, so we split LMA into nitrogen-based protein and other CBC contents, following:

$$LMA = Protein\ content + CBC\ content.$$

LMA, protein content and CBC content are expressed in mass per leaf surface unit (mg.cm$^{-2}$), which ensures conservation of the mass of absorbing materials and allows us to obtain LMA directly as the sum of leaf protein and CBC content estimates.

Absorption $k(\lambda)$ of a compact leaf layer at wavelength $\lambda$, characterized by a mesophyll structural parameter $N_{struct}$, is in the PROSPECT model defined as:

$$k(\lambda) = \frac{\sum_i K_{spe,i}(\lambda) \times C_i}{N_{struct}},$$

where $K_{spe,i}(\lambda)$ is the SAC of a constituent $i$, and $C_i$ is its corresponding content. Our study focuses on the spectral region from 1000 nm to 2500 nm. In PROSPECT-D, only two input constituents contribute to absorption in these wavelengths: water (EWT), with a negligible absorption before 1100 nm, and dry matter (LMA) with a constant absorption between 1000 and 1200 nm. The brown pigments were excluded from our analysis, as they have a low level of absorption between 1000 and 1100 nm. Therefore, Eq. 3 can be for the purpose of PROSPECT calibration between 1000 and 2500 nm written as follows:

$$k(\lambda) = \frac{K_{spe,EWT}(\lambda) \times C_{EWT} + K_{spe,LMA}(\lambda) \times C_{LMA}}{N_{struct}}.$$

Following the principle in Eq. 2, we defined constraints for the specific absorption coefficients corresponding to leaf protein and CBC content. The contribution of LMA to the total absorption can then be decomposed into the proteins and CBCs as:

$$K_{spe,LMA}(\lambda) \times C_{LMA} = K_{spe,PROT}(\lambda) \times C_{PROT} + K_{spe,CBC}(\lambda) \times C_{CBC},$$ Eq. 5

where $K_{spe,PROT}(\lambda)$ is SAC for proteins, $K_{spe,CBC}(\lambda)$ is SAC for the CBC (both in cm².mg⁻¹), and $C_{PROT}$ and $C_{CBC}$ are the corresponding contents (in mg.cm⁻²), respectively.

The calibration process of PROSPECT-PRO followed the two-step process described in Féret et al. (2008, 2017). First, we determined the leaf structure parameter $N_{struct,j}$ of each leaf $j$ in the calibration datasets. $N_{struct,j}$ was estimated based on a multivariate iterative optimization, simultaneously with three absorption coefficients, using reflectance and transmittance values measured at three wavelengths corresponding to the minimum absorptance ($\lambda_1$), maximum reflectance ($\lambda_2$), and maximum transmittance ($\lambda_3$) of the leaf (Jacquemoud et al., 1996). These values are generally located on the NIR reflectance and transmittance plateau. The iterative optimization was performed using the following merit function:

$$M_{leafN}\left(N_{struct,j}, k(\lambda_1), k(\lambda_2), k(\lambda_3)\right) = \sum_{l=1}^{3}\left[\left(R_{meas,j}(\lambda_l) - R_{mod}\left(N_j, k(\lambda_l)\right)\right)^2 + \right.$$
$$\left.\left(T_{meas,j}(\lambda_l) - T_{mod}\left(N_j, k(\lambda_l)\right)\right)^2\right],$$ Eq. 6

where $R_{meas,j}(\lambda_l)$ and $T_{meas,j}(\lambda_l)$ are measured directional-hemispherical reflectance and transmittance of leaf $j$ at wavelength $\lambda_l$, $R_{mod}$ and $T_{mod}$ are the respective modeled values, and $k(\lambda)$ is the specific absorption coefficient of a compact layer at the wavelength $\lambda$, which is being adjusted simultaneously with $N_{struct,j}$.

In the second step, $K_{spe,PROT}$ and $K_{spe,CBC}$ were computed by inverting PROSPECT-PRO using the CALIBRATION dataset for each spectral band of interest independently. We minimized the following merit function $J$ per wavelength:

$$J\left(\{K_{spe,i}(\lambda)\}_{i=1:n}\right)$$
$$= \sum_{j=1}^{n}\left[\left(R_{meas,j}(\lambda) - R_{mod,j}\left(N_{struct,j}, k(\lambda)\right)\right)^2\right.$$
$$\left. + \left(T_{meas,j}(\lambda) - T_{mod,j}\left(N_{struct,j}, k(\lambda)\right)\right)^2\right],$$

Eq. 7

with $k(\lambda)$ as defined in Eq. 3. Following the constraints defined in Eq. 2 and Eq. 5, $k(\lambda)$ was expressed as:

$$k(\lambda) = \frac{K_{spe,EWT}(\lambda) \times C_{EWT} + K_{spe,PROT}(\lambda) \times C_{PROT}}{N_{struct}}$$
$$+ \frac{\frac{\left(K_{spe,LMA}(\lambda) \times C_{LMA} - K_{spe,PROT}(\lambda) \times C_{PROT}\right)}{C_{LMA} - C_{PROT}} \times (C_{LMA} - C_{PROT})}{N_{struct}},$$

Eq. 8

where $K_{spe,PROT}$ is the only unknown parameter adjustable through the iterative optimization.

The calibration was performed within the spectral domain ranging from 1000 nm to 2500 nm, using the same leaf refractive index and SAC for EWT as defined in PROSPECT-D.

c. Global sensitivity analysis of PROSPECT-PRO

A global sensitivity analysis (GSA) was carried out for PROSPECT-PRO to quantify the contribution of proteins and CBC constituents to the overall spectral signal. Using a GSA the driving variables of a radiative transfer model can be identified by fully exploring the input parameter space (Verrelst et al., 2019b; Wang et al., 2015). The Matlab software tool GSAT (Cannavó, 2012), which includes Fourier amplitude sensitivity testing (FAST) analysis and Sobol's method for calculation of the first-order sensitivity coefficients was applied on PROSPECT-PRO simulations from 1000 nm to 2500 nm carried out with the following realistic input parameter ranges for fresh leaves: $N_{struct}$ ~ 1-2 (unitless), EWT ~ 0.001-0.015 cm, protein content ($C_p$) ~ 0-0.003 g/cm² and CBC content ~ 0-0.01 g/cm². The remaining PROSPECT-PRO input parameters, i.e. chlorophyll content, total carotenoid content, anthocyanin content and brown pigment content, were fixed to arbitrary values since they manifest no absorption between 1000-2500 nm.

d.  Optimal spectral domains for estimation of protein and CBC content

PROSPECT-PRO inversion was based on iterative optimization to identify the best combination of leaf biochemical and structural parameters to minimize following merit function:

$$M_{inv}(N_{struct}, \{ C_i \}_{i=1:p}) = \sum_{\lambda=1}^{n_\lambda} \Big[ \big(R_{meas,\lambda} - R_{mod,\lambda}(N_{struct}, \{ C_i \}_{i=1:p})\big)^2 + \big(T_{meas,\lambda} - T_{mod,\lambda}(N_{struct}, \{ C_i \}_{i=1:p})\big)^2 \Big],$$

Eq. 9

where $n_\lambda$ is the number of available spectral bands, $N_{struct}$ is the leaf inner structure parameter, $C_i$ is the content of constituent $i$, and $p$ is the number of leaf biochemical constituents. In many studies, a PROSPECT model inversion was performed using the full spectral information or specific domains, such as the VIS-NIR when retrieving leaf pigments and NIR-SWIR domain when retrieving EWT or LMA.

Féret et al. (2019) showed the importance of identifying the optimal spectral domain for the accurate estimation of LMA and, to a lesser extent, EWT. Investigating a number of spectral ranges between 1000 nm to 2400 nm, they recommended determining EWT and LMA using an iterative optimization of leaf optical properties between 1700 and 2400 nm. As proteins contribute to LMA, the hypothesis that proteins and LMA share the same optimal spectral domain for their retrieval is in need of testing. However, protein absorption is expressed in narrow absorption features in the SWIR (Curran, 1989; Fourty et al., 1996), which suggests that proteins and CBCs may have slightly different optimal retrieval subdomains compared to LMA. Therefore, we applied a procedure similar to Féret et al. (2019) to identify optimal spectral subdomains for retrieval of proteins and CBCs. We compared the performances of three inversion strategies sharing the same merit function but using different PROSPECT versions and spectral domains:

1. Inversion of PROSPECT-D using LOP from 1700 nm to 2400 nm to estimate simultaneously $N_{struct}$, EWT and LMA.

2. Inversion of PROSPECT-PRO using LOP from 1700 nm to 2400 nm to estimate simultaneously $N_{struct}$, EWT, protein and CBC contents. LMA was estimated by following Eq. 2.

3. Inversion of PROSPECT-PRO using LOP with a spectral subdomain identified as optimal for either protein or CBC content to estimate simultaneously $N_{struct}$, EWT, protein and CBC content. LMA was then estimated by following Eq. 2.

We assume that the optimal retrieval subdomains for proteins or CBCs may correspond to narrow spectral intervals that do not include relevant information for estimation of other constituents, in particular EWT. Therefore, our object was not to identify the optimal subdomain for the simultaneous estimation of all constituent contributing to SWIR absorption, but rather for each separately. Table 3 summarizes the retrieved leaf constituents to be compared among inversion strategies. The comparison between individual estimation of proteins and CBC content was performed with the VALIDATION dataset only, while the comparison between estimation of EWT and LMA (either directly estimated with Inversion #1 or as a sum of estimated protein and CBC content with Inversion #2 and #3) was performed with the additional six datasets (Table 2).

Table 3. Overview of PROSPECT retrieved leaf constituents to be compared among the different inversion strategies tested in this study.

| Inversion #1 | Inversion #2 | Inversion #3 |
|---|---|---|
| EWT | EWT | *N/A* |
| LMA | Proteins + CBC | Proteins + CBC |
| *N/A* | Proteins | Proteins |
| *N/A* | CBC | CBC |

The normalized root means square error (*NRMSE* expressed in %) was computed to appraise the difference between the measured and estimated leaf constituents retrieved from the different datasets:

$$NRMSE = \frac{1}{\overline{X_{meas}}} \sqrt{\frac{\sum_{j=1}^{n}\left(X_{meas,j} - X_{mod,j}\right)^2}{n}}, \qquad \text{Eq. 10}$$

where $X_{meas,j}$ is the measured value and $X_{mod,j}$ is the values estimated by model inversion for a leaf $j$, $\overline{X_{meas}}$ is the mean value of the constituent, and $n$ is the number of samples.

Féret et al. (2019) identified the optimal estimation spectral domain for LMA and EWT by dividing the NIR-SWIR from 1000 to 2399 nm into 100 nm segments, and testing PROSPECT-D inversion using all combinations of segments that produced contiguous spectral subdomains. Using this same strategy for leaf protein and CBC content, for PROSPECT-PRO we tested all contiguous combinations of 40 evenly sized segments of 25 nm each between 1400-2399 nm (n = $(41 \times 40)/2 = 820$), the spectral domain covering the main absorption features associated with proteins (Curran, 1989; Fourty et al., 1996). We considered the subdomain that minimized the RMSE for the independent estimation of protein and CBC content from the VALIDATION dataset as optimal.

e.  Performances of PROSPECT-D and PROSPECT-PRO in forward modelling

We also compared the performances of PROSPECT-D and PROSPECT-PRO for the forward simulation of leaf optical properties, using the structure parameter $N_{struct}$ obtained from inversion and the corresponding biochemical constituents measured in laboratory. This statistical analysis was undertaken to reveal spectral domains impacted by high levels of uncertainty, which is relevant for hybrid inversion applications involving machine learning algorithms trained with PROSPECT-PRO simulated data (e.g., Verrelst et al., 2015 and 2019a). The comparison was performed by computing the per-wavelength spectral RMSE between measured and simulated reflectance and transmittance of fresh and dry samples from the VALIDATION dataset.

f.  Estimation of the C:N ratio

The C:N ratio of plant canopies, crops and crop residues is of great importance for modelling C and N dynamics in natural ecosystem and agricultural systems, as it contains an indicative information about plant growth rate and affects ecosystem response to $CO_2$ (Reich et al., 2006; Zheng, 2009). This C:N ratio is also an indicator of the relative allocation of resources in vegetation, an indicator of potential decomposition rate of litter and an important factor promoting soil organic carbon accumulation (Zhou et al., 2019). Thus, we tested the possibility of using the CBCs:Proteins ratio, estimated from PROSPECT-PRO inversion, as a proxy for the C:N ratio of leaf samples in the LOPEX dataset. We established a linear model to estimate the C:N ratio based on the CBCs:Proteins ratio as measured in the fresh samples of the CALIBRATION dataset. We then applied this linear relationship on CBC:Proteins ratio retrieved from LOP through PROSPECT-PRO inversion to estimate the C:N ratio for all samples in both CALIBRATION and VALIDATION datasets.

5. RESULTS

    a. Correlations among biochemical constituents of the fresh leaves from LOPEX-CALVAL

The correlation analysis performed on the leaf constituents of fresh samples in LOPEX-CALVAL was aiming to evidence potential relationships between individual biochemical compounds. Figure 1 shows the Pearson correlation coefficients calculated among the constituents, including also the C:N ratio. Coefficients highlight strong and statistically significant relationships between carbon, hydrogen, oxygen, lignin, cellulose and LMA. Proteins are not included in the figure, as they were directly derived from N measurements. N is moderately correlated to CHL, C, H, O, LMA and EWT. The moderate correlation between CHL and N (r=0.51) indicates the modest capacity of chlorophyll a+b content to estimate nitrogen across species in the LOPEX dataset. LOPEX includes species found in a unique region, which means that this correlation analysis may not lead to the same conclusion as analyses performed over broader vegetation types. Finally, the C:N ratio was found to be positively correlated with LMA, C. H. O and individual CBCs, and poorly correlated with N content.

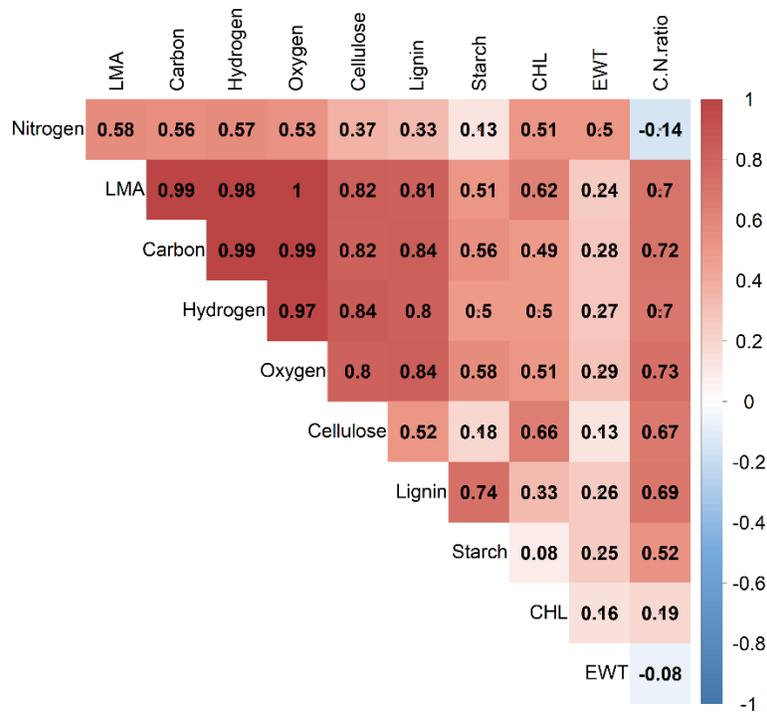

Figure 1. Correlation among leaf constituents in the fresh samples of the LOPEX-CALVAL dataset.

b. Calibration of PROSPECT-PRO

Once the SAC for proteins and CBCs were derived, they were evaluated to ensure the compatibility between PROSPECT-PRO and PROSPECT-D and to prevent incongruities in spectral regions with otherwise low absorption (e.g. the NIR). The retrieved SAC corresponding to proteins showed a negligible absorption between 1200 and 1500 nm. Because of higher uncertainties in both modelled and experimental data in the NIR domain (Féret et al., 2019), SAC values were adjusted to 0 for wavelengths < 1440 nm and an exponential model was applied between 1440 nm and 1484 nm to smooth transition between the non-absorptive and absorptive spectral domains. Subsequently, the SAC values for CBC content were adjusted according to the SAC of LMA from PROSPECT-D by applying a multiplicative factor corresponding to the average ratio between LMA and CBC in the CALIBRATION dataset. The resulting SACs of leaf protein and CBC are displayed in Figure 2. Most of the absorption features reported by Curran (1989) and Fourty et al. (1996) correspond with local maxima in the SAC of proteins, although some of them spectrally shifted towards shorter or longer wavelengths. A Matlab version of the PROSPECT-PRO model is downloadable from the following GitLab repository: https://gitlab.com/jbferet/prospect_pro_matlab.

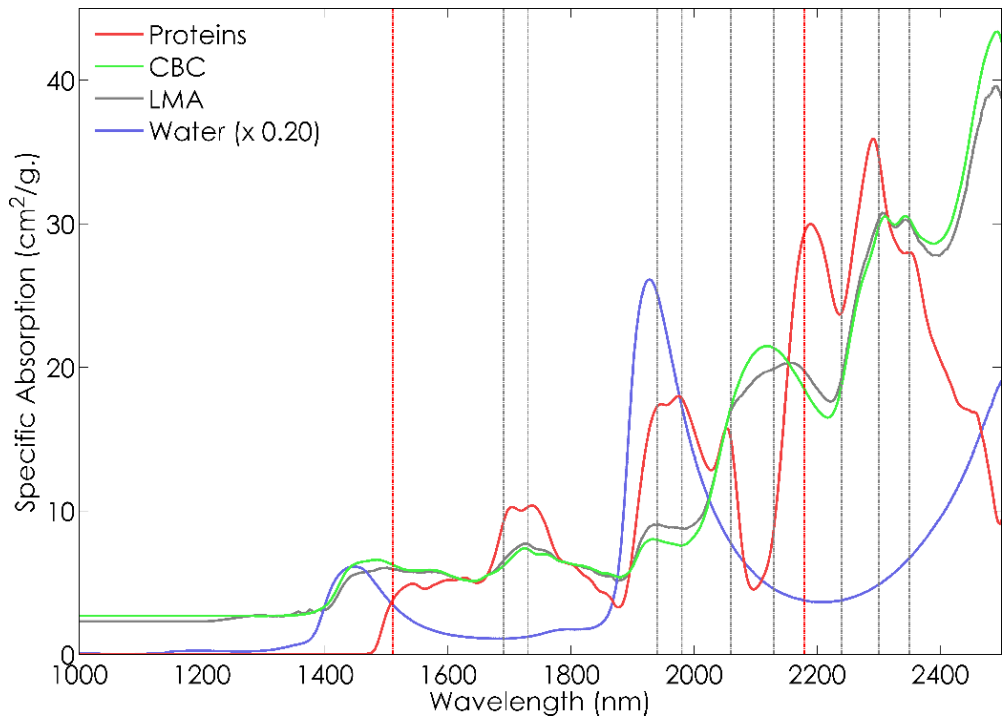

Figure 2. Specific absorption coefficients (SACs) for proteins and CBC obtained from calibration of

PROSPECT-PRO using the CALIBRATION dataset. The SAC corresponding to LMA and EWT (Water;

scale x 0.20) in PROSPECT-D are displayed for comparison and to identify main absorption features.

Vertical lines correspond to absorption features linked to proteins reported by Curran (1989) and

Fourty et al. (1996) (red = major absorption features).

c. Sensitivity of leaf optical properties to proteins and CBC

Results of GSA (Figure 3) identify the spectral regions expected to have to absorption peaks of proteins

(>1400 nm), but also shows the low contribution of proteins to the overall spectral signal in these wavelengths. CBCs play a larger role in driving leaf reflectance, with their highest relevance in the SWIR, especially above 2000 nm. Moreover, the key driving input parameters of PROSPECT-PRO forward simulations are the $N_{struct}$ parameter and EWT. The subtle but still present absorption features of proteins between 1600 and 1800 nm and between 2100 and 2300 nm confirm that these spectral domains are likely most suitable for retrieval activities. However, a high signal-to-noise (SNR) is required to enable the separation of all influencing constituents, especially in future efforts when upscaling the retrieval methods from the leaf to the top-of-canopy level.

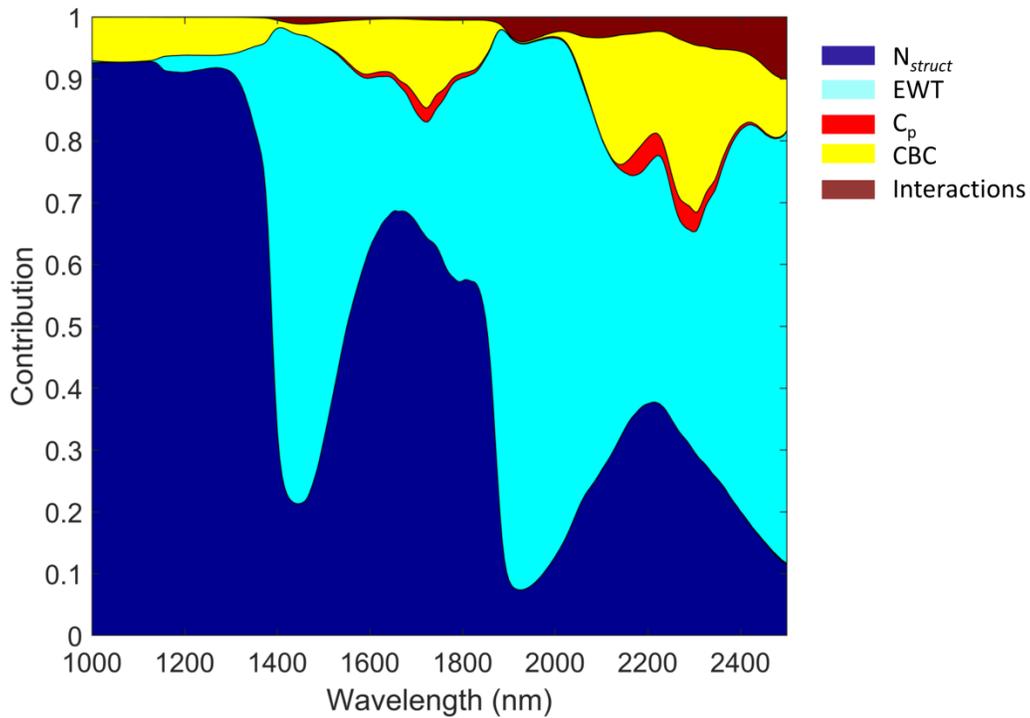

Figure 3. Global sensitivity analysis of PROPECT-PRO with interactions and influence of input parameters corresponding to leaf structure ($N_{struct}$), leaf water content (EWT), protein content ($C_p$) and carbon-based constituents (CBC) of a typical leaf. The y-axis ('contribution') quantifies the main (first-order) effects, implying the contribution of each tested input to the modelled output variance.

d. Optimal spectral domains for PROSPECT-PRO retrieval of proteins and CBC content

Figure 4 shows the performance for the estimation of protein and CBC content by inverting PROSPECT-PRO using the VALIDATION dataset when testing all 25 nm or greater contiguous domains between 1400 and 2399 nm. The performances are expressed as standardized RMSE (sRMSE), where the RMSE obtained with Inversion #2 (using leaf optical properties between 1700 and 2400 nm) is considered to be the standard value with a baseline of 100%.

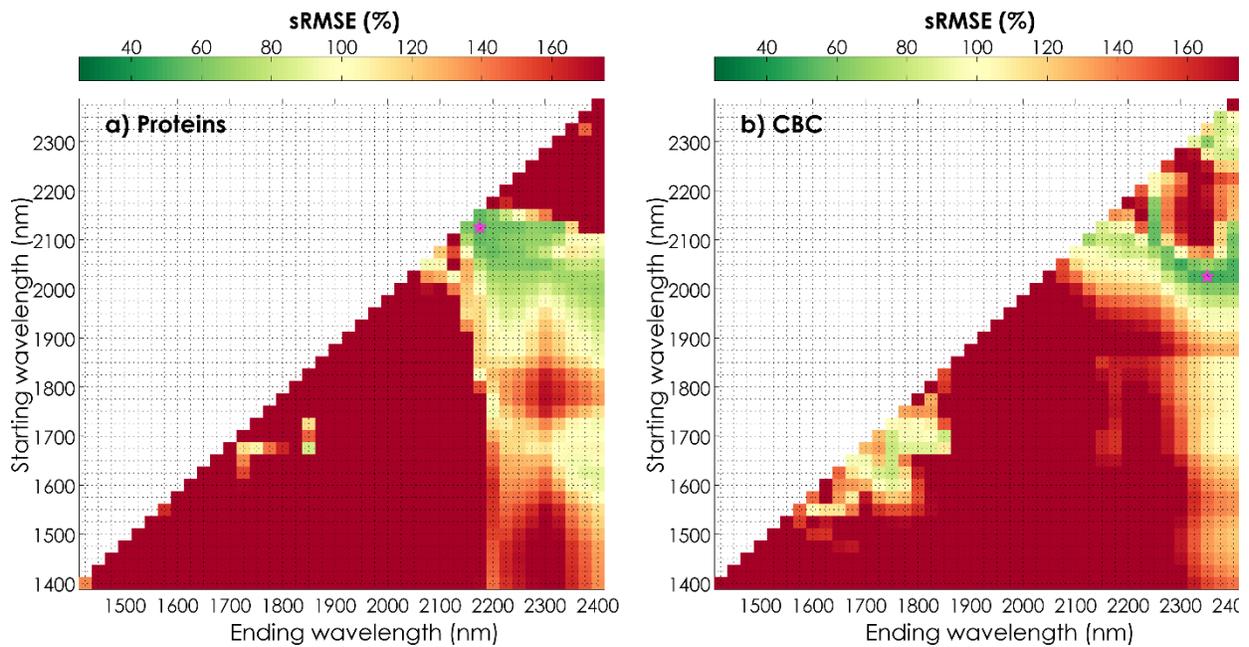

Figure 4. sRMSE (%) obtained for the estimation of a) proteins and b) CBC content with PROSPECT-PRO inversion applied on the VALIDATION dataset using reduced spectral domains bounded by a starting wavelength λ1 (y-axis) and an ending wavelength λ2 (x-axis). The standardization baseline is the performance obtained when using the entire domain from 1700 nm to 2400 nm (sRMSE = 100%). The pink star indicates spectral intervals with the lowest sRMSE.

The optimal spectral domain for the estimation of leaf protein content was found between 2125 and 2174 nm. This spectral domain is located next to the strong absorption feature of proteins centered at 2180 nm noted by Curran (1989), Fourty et al. (1996) and Wang et al. (2015). Use of spectral information between 1400 and 1600 nm strongly decreased the capability to estimate protein content. The optimal spectral domain for leaf CBC content was 2025 to 2349 nm. The broader interval obtained when estimating leaf CBCs compared to leaf protein content includes spectral absorption features corresponding to lignin, cellulose, starch and sugars.

e. PROSPECT-PRO validation by retrieval of leaf protein and CBC contents

Here we compare PROSPECT-PRO performances obtained for estimation of leaf protein and CBC content using Inversion #2 (simultaneous estimation of leaf proteins and CBC using the spectral information from 1700 nm to 2400 nm) to Inversion #3 (estimation of leaf proteins and CBCs using the optimal spectral domains). Inversion #3 outperformed Inversion #2 (Table 4).

Table 4. NRMSE (%) and $R^2$ computed between laboratory measured and PROSPECT-PRO estimated leaf protein and CBC content for fresh and dry samples of the VALIDATION dataset, obtained either from Inversion #2 or from Inversion #3.

| Inversion | Inversion#2 | | Inversion#3 | |
|---|---|---|---|---|
| Spectral domain | 1700-2400 | | OPTIMAL | |
| | NRMSE (%) | $R^2$ | NRMSE (%) | $R^2$ |
| Proteins | | | 2125-2174 nm | |
| **Dry samples** | 33.7 | 0.45 | 24.0 | 0.62 |
| **Fresh samples** | 47.1 | 0.55 | 17.3 | 0.75 |
| CBC | | | 2025-2349 nm | |
| **Dry samples** | 14.4 | 0.94 | 13.9 | 0.95 |
| **Fresh samples** | 37.2 | 0.86 | 13.1 | 0.92 |

Comparison of estimated and measured leaf protein, CBC content and LMA (as the sum of proteins and CBC content) for Inversions #2 and #3 illustrates the stronger performance of the optimal domain (Figure 5). The estimation of proteins has slightly higher uncertainty than the estimation of CBCs and LMA, which was expected given the lower contribution of proteins to the spectral signal (Figure 3). Figure 5 also explains the lower performances obtained for the estimation of protein content from dry leaves reported in Table 4, caused mainly by a single sample.

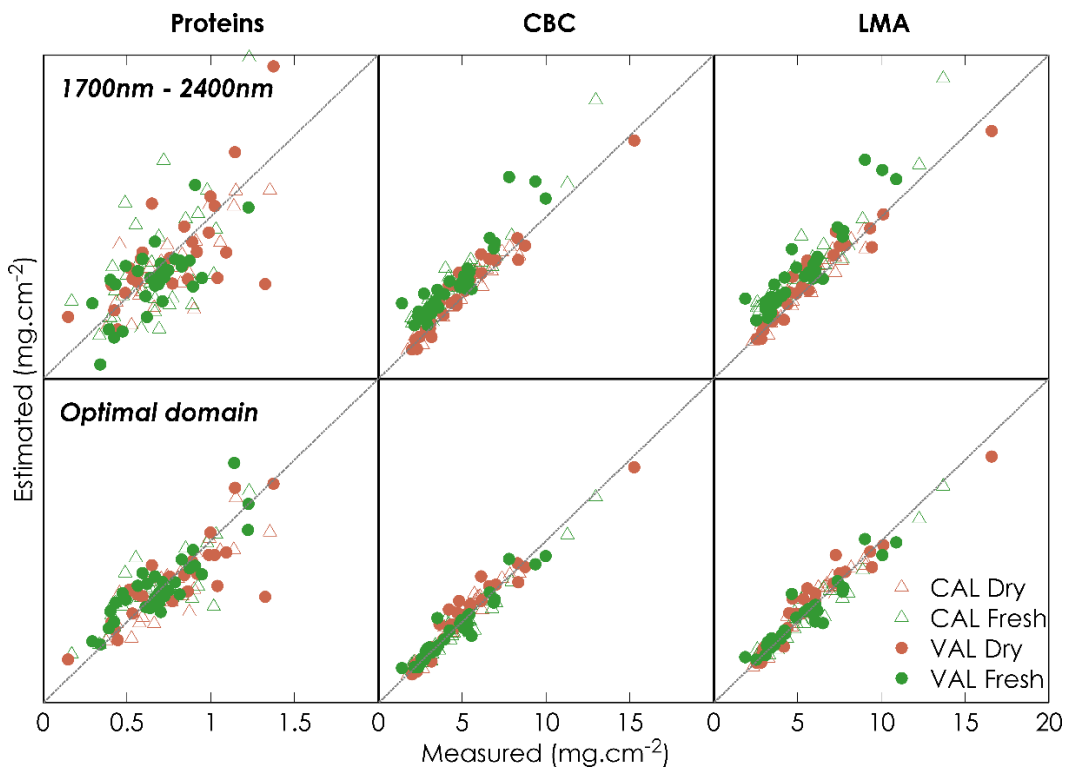

Figure 5. Comparison between laboratory measured and PROSPECT-PRO estimated leaf protein content, CBC content and LMA (protein + CBC) obtained either from Inversion #2 or from Inversion #3, for dry and fresh samples in the CALIBRATION dataset (brown and green triangles), and for dry and fresh samples in the VALIDATION dataset (brown and green dots).

f.   PROSPECT-PRO and PROSPECT-D compatibility assessed via estimation of LMA and EWT

Overall, the decomposition of LMA into protein and CBC contents estimated by the PROSPECT-PRO Inversion #3 slightly outperformed the estimation of LMA obtained with the PROSPECT-D inversion on the optimal spectral domains that were found by Féret et al. (2019) (

Table 5, Figure 6). When analyzing the results per dataset, inversion of PROSPECT-PRO using the optimal spectral domain for protein and CBC content resulted in a decreased NRMSE for six of eight datasets and a moderate NRMSE increase for the other two datasets, with a 2.3% decrease in NRMSE and comparable $R^2$ across all data sets. The increase in NRMSE observed for ANGERS is mainly due to two samples with very high LMA, while the remaining samples showed lower NRMSE. Finally, the version of the model did not influence the estimation of EWT significantly, as the Inversion #1 and #2 resulted in similar outcomes (results not shown). These results on EWT and LMA retrievals confirm the compatibility between PROSPECT-D and PROSPECT-PRO.

Table 5. NRMSE (%) for estimations of LMA retrieved by PROSPECT-D and PROSPECT-PRO with two inversion strategies applied on different datasets, including those independent from calibration. Bold values correspond to best performances for a given dataset.

| Model | PROSPECT-D | | PROSPECT-PRO | | | |
|---|---|---|---|---|---|---|
| Spectral domain | 1700-2400 | | 1700-2400 | | Optimal domain | |
| Criterion/dataset | NRMSE (%) | $R^2$ | NRMSE (%) | $R^2$ | NRMSE (%) | $R^2$ |
| *VAL. (Dry)* | 14.2 | 0.93 | 14.1 | **0.94** | **13.5** | **0.94** |
| *VAL. (Fresh)* | 34.3 | 0.83 | 33.0 | 0.87 | **12.8** | **0.91** |
| *ANGERS* | 18.1 | 0.95 | **17.8** | **0.95** | 21.2 | 0.94 |
| *HYYTIALA* | **24.3** | 0.79 | 25.0 | **0.80** | 25.0 | 0.78 |
| *ITATINGA* | 14.0 | 0.68 | 14.7 | 0.68 | **10.6** | **0.74** |

| | | | | | |
|---|---|---|---|---|---|
| ***NOURAGUES*** | 14.6 | **0.89** | 17.2 | **0.89** | **14.5** | 0.86 |
| ***PARACOU*** | 14.1 | **0.85** | 15.9 | **0.85** | **13.5** | 0.83 |
| **LOPEX-Full** | 37.5 | **0.80** | 35.2 | 0.75 | **29.2** | 0.71 |
| **Combined** | 18.6 | **0.91** | 19.5 | **0.91** | **16.3** | 0.90 |

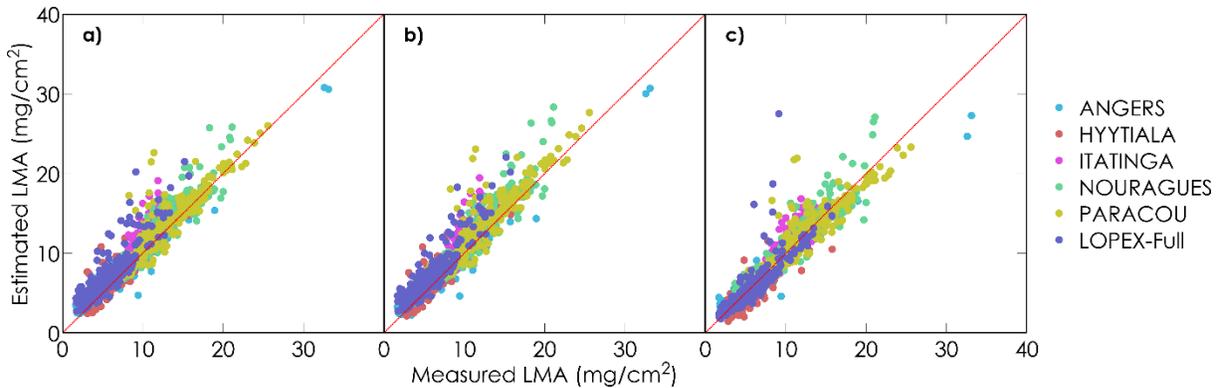

Figure 6. Comparison between measured LMA and its corresponding estimations by a) Inversion #1 (PROSPECT-D), b) Inversion #2 (PROSPECT-PRO using 1700-2400 nm), and c) Inversion #3 (PROSPECT-PRO using optimal spectral domains for leaf proteins and CBC content).

g. Forward simulations of leaf optical properties

Figure 7 displays per-wavelength RMSE calculated between measured leaf reflectance and transmittance and their counterparts simulated by PROSPECT-PRO for dry and fresh samples of the VALIDATION dataset as well as the six independent datasets reported in Table 2 together. The values of chemical constituents used as input variables for the VALIDATION samples correspond to the values obtained from laboratory measurements, and the $N_{struct}$ parameter was obtained from the model Inversion #2 (for PROSPECT-D) or #3 (for PROSPECT-PRO), respectively. Since the independent datasets (Figure 7c) do not contain protein and CBC content measurements, the forward modelling for those was carried using PROSPECT-D. In this case, the results corresponding to the RMSE between LOP measurements and PROSPECT-PRO simulations are based on values of protein and CBC contents obtained from Inversion #3. As well, the PARACOU dataset was excluded from this analysis, as no measurements of EWT were available.

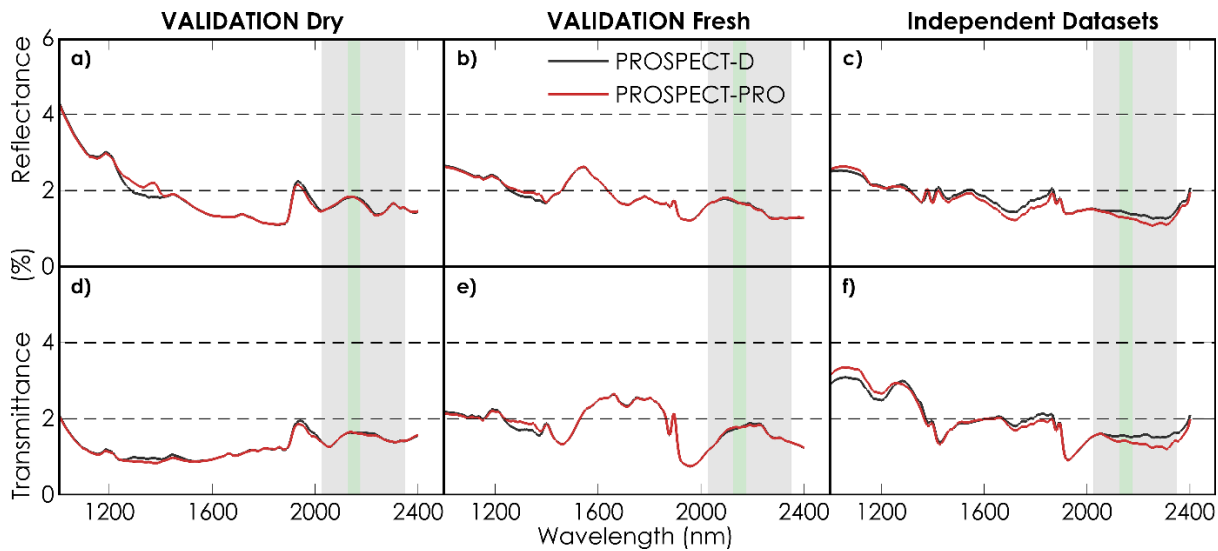

Figure 7. RMSE between measured and simulated LOPs obtained from PROSPECT-D and PROSPECT-PRO forward modelling. The chemical constituents were measured in laboratory and the $N_{struct}$ parameter was derived from each corresponding PROSPECT inversion using the spectral domain from 1700 nm to 2400 nm. The green and grey areas highlight the respective optimal spectral domains identified in this study for estimation of protein and CBC content.

The dry samples from VALIDATION exhibited an RMSE between 1 and 2.2% in the SWIR, increasing in the NIR to as high as 4% for reflectance and 2% for transmittance at 1000 nm. The increasing RMSE at shorter wavelengths may be due to the presence of constituents similar to brown pigments appearing after the drying process, which were unaccounted during the simulation, or by residual model inaccuracies at these wavelengths. The results obtained for dry samples show that the SAC of LMA as well as proteins and CBC constituents match well LOPEX data, except for the NIR domain between 1000 nm and 1200 nm. This was partially expected, as the LOPEX data was used by Féret et al. (2008) to calibrate the LMA SAC of PROSPECT-D .

In case of fresh VALIDATION samples, the RMSE is slightly higher and fluctuates between 1 and 3% of reflectance and transmittance intensities, but the RMSE in the optimal domain for the estimation of proteins and CBC is between 1 and 2%. The lower RMSE between 1000 and 1200 nm compared to dry samples suggests supports our interpretation of increased NIR RMSE in dry samples due to the presence of absorbing constituents after the drying process.

For the independent datasets, RMSE lower than 2% was found for both reflectance and transmittance for wavelengths > 1500 nm and between 1 and 2% for wavelengths > 2000 nm, which is in agreement with the results obtained for the fresh samples from VALIDATION. The RMSE between experimental LOP and simulations obtained with PROSPECT-PRO with the protein and CBC derived from Inversion #3 show very similar values to those obtained with PROSPECT-D.

We conclude that PROSPECT-D and PROSPECT-PRO are compatible, as they exhibit similar performance in forward simulations of LOPs. Both PROSPECT-PRO and PROSPECT-D have moderate overall error between measured and simulated LOPs in the NIR and SWIR domains (between 1 and 3%).

h.   Estimation of C:N from CBC:Proteins ratio retrieved from PROSPECT-PRO inversion

The correlation analysis displayed in Figure 1 shows that constituents of CBC, such as cellulose and lignin, are strongly correlated with leaf C content. We applied the linear model fitted between CBC:Proteins and C:N ratio of the fresh samples in the CALIBRATION dataset (Eq. 11) to the CBC:Proteins ratio PROSPECT-PRO estimates for both CALIBRATION and VALIDATION datasets:

$$C:N = 2.157 \times CBC:Prot + 1.515.$$
<div align="right">Eq. 11</div>

The results show that the C:N ratio was derived from the CBC and protein contents estimated through the PROSPECT-PRO inversion with a relative accuracy between 68.8 and 87.3% (Table 6). Measured and estimated C:N ratios aligned well (Figure 8). Overall, the C:N ratio was estimated with the best accuracy for the fresh VALIDATION samples. The poorer performances for fresh CALIBRATION and dry VALIDATION datasets were strongly driven by a single sample, which highlights the importance of further independent validation of the regression model.

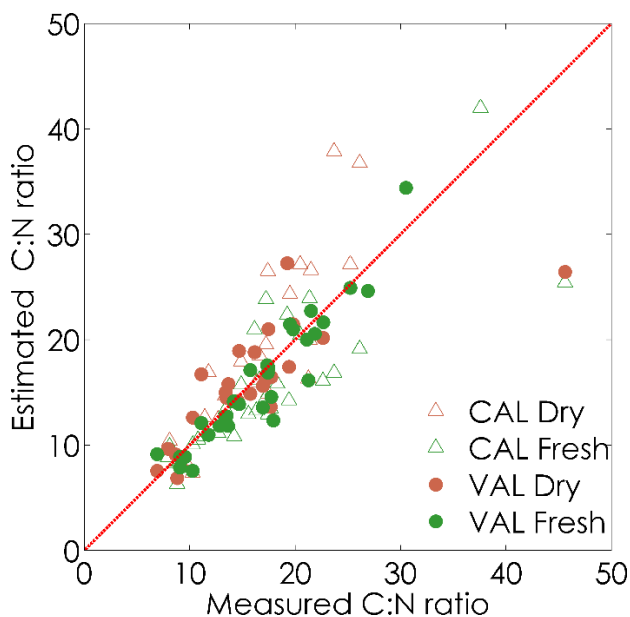

Figure 8. Comparison between the C:N ratio measured in laboratory and the same ratio derived from regression (Eq. 11) with protein and CBC contents estimated through the PROSPECT-PRO inversion of dry and fresh samples in the CALIBRATION and VALIDATION datasets.

Table 6. $R^2$ and NRMSE (%) computed for the estimation of the C:N ratio for samples of CALIBRATION and VALIDATION datasets, based on the CBC:Proteins ratio estimated from PROSPECT-PRO inversion.

| Dataset | Samples | $R^2$ | NRMSE (%) |
|---|---|---|---|
| CALIBRATION | Dry | 0.78 | 31.2 |
| | Fresh | 0.66 | 29.0 |
| VALIDATION | Dry | 0.58 | 30.7 |
| | Fresh | 0.89 | 12.7 |

6. DISCUSSION

    a. Contribution towards operational remote sensing retrieval methods

Physical RT modeling is a key component in revealing the underlying relationships between quantitative vegetation properties and information encoded in RS optical data. In our study, we introduce a new PROSPECT-PRO model that separates nitrogen-based protein constituents from carbon-based constituents, i.e. cellulose, lignin, hemicellulose and starch. PROSPECT-PRO was successfully calibrated and validated using measurements of the LOPEX dataset. Unlike previous attempts to incorporate proteins in the model for fresh leaves, which either failed (Jacquemoud et al., 1996) or showed limited accuracy in LMA predictions (Wang et al., 2015), the indirect estimation of LMA with PROSPECT-PRO, computed as the sum of proteins and CBC, revealed comparable performance with the direct estimation of LMA using PROSPECT-D, although its performance at high levels of LMA (> 20 mg cm$^{-2}$) may be poor, based on VALIDATION data. The performances for forward simulations of LOPs were also very similar between the two versions, demonstrating their full compatibility. This ensures modelling consistency across the PROSPECT versions, while opening up a new functionality of estimating protein and separating its contribution to LMA from that of CBCs, which subsequently enables estimation of nitrogen relative content. Similar to Féret et al. (2019), our results illustrate that an accurate estimation of LMA and its two components depends on selection of the appropriate spectral domain for the inversion. The negative influence of water SWIR absorption on the retrieval of dry matter constituents must be accounted for in the future operational applications for nitrogen or LMA monitoring using field as well as air-/space-borne imaging spectroscopy. This limitation may be alleviated by applying appropriate weights for the different spectral domains to optimize the sensitivity of retrieval algorithms to constituents of interest. The most unsuitable spectral wavelengths can be identified and removed by hybrid methods featuring band selection, feature extraction or band weighting procedures (Fassnacht et al., 2014; Feilhauer et al., 2015; Verrelst et al., 2015). Unfortunately, the residual inaccuracy of PROSPECT-PRO in the NIR domain cannot be excluded from the pool of potential errors and needs to be accounted for.

b. Potential uncertainty in the experimental data

Uncertainties are related not only to the physical-empirical model design and mathematical inversion but also to model inputs, i.e. leaf biochemical and optical measurements (Malenovský et al., 2019). We used a single nitrogen-to-protein content conversion multiplicative factor of 4.43 (Yeoh and Wee, 1994). Yet, this factor is not constant across all plant species. Various authors report that it can range from 3.28 to 5.16, with an average and standard deviation of 4.43±0.40. This means that the protein content used for calibration and validation of PROSPECT-PRO contains an associated uncertainty that is proportional to the unaccounted variability of this conversion factor. This may also explain the moderately higher uncertainty observed for protein estimates when compared to LMA and CBC. Despite this uncertainty, our results show that the SAC for in vivo proteins are consistent with SAC derived from dried and ground leaves reported in literature (Curran, 1989). In addition, the protein content estimated through model inversion remained consistent and accurate.

Our study only includes one dataset with measured protein and CBC content. Therefore, the validation is performed on a limited number of samples (n=26 dry and 33 fresh). As such, the errors and uncertainties reported might be strongly affected by just few discrepancies in this low number of samples. As reported in Section 3.a, the presence or absence of a single sample in the calibration dataset significantly impacted the calibration process and the subsequent capability of the model to properly estimate leaf constituents and simulate LOP. In the same way, the presence of a limited number of samples showing strong error may lead to difficulties for the statistical analysis of results obtained after inversion. In our case, the lower performance obtained for the estimation of proteins on dry samples, when compared to fresh samples, was caused by a unique sample, for which measurement error may have occurred. More public datasets of reliable full leaf optical properties and corresponding comprehensive and robust laboratory measurements of leaf constituents are strongly needed for future improvements of PROSPECT, as well as to better identify the limitations of PROSPECT inversion (for high EWT and LMA as identified for a limited set of samples in the current study) and to explore possibilities to differentiate new constituents such as structural and nonstructural carbohydrates (Durán et al., 2019).

c.  Complementarity of chlorophyll and protein estimates as proxies for nitrogen content

Despite the known limitations of using chlorophyll a+b content as a proxy of N in remote sensing monitoring applications, it has proved to be relatively successful in a certain number of cases (Baret et al., 2007). The main advantage of estimating chlorophyll over protein content is its strong optical signal in the VNIR (especially red-edge) domain, allowing for its accurate RS estimates even at the canopy level (e.g. Malenovský et al., 2013). In contrast, the SWIR domain is required to estimate protein content, but is characterized by lower solar energy flux and consequently lower SNR (Guanter et al., 2015). Therefore, even if being physiologically more robust over a broader range of conditions and vegetation types, the estimation of protein content, and subsequently N, may be associated with a significantly higher uncertainty due to a weaker SNR of the spectroscopic measurements. The enhanced capacity of PROSPECT-PRO to monitor vegetation C:N ratio and its seasonal changes through the separation between protein and CBC content may prove useful if systematically and rationally complemented by a RS chlorophyll monitoring.

d.  Potential application for a canopy scale ecosystem nitrogen mapping

Our results offer a new perspective in operational RS monitoring and consequent management of nitrogen in agricultural and natural ecosystems. However, these applications are strongly dependent on being able to also scale leaf level results up to spatially and spectrally heterogeneous canopies. The potential for transferring our methods to operational applications in the field is yet to be investigated. Close-range remote sensing applications, based for instance on the PROCOSINE model (Jay et al., 2016; Morel et al., 2018), could be considered as an intermediate scale between leaf and canopies, but a certain number of challenges would need to be addressed first. These include the capacity to perform outdoor canopy measurements with an acceptable uncertainty induced by the vegetation canopy reflectance angular anisotropy (including background surfaces and leaf orientation) in combination with the need for sufficient SNR in the SWIR domain. Based on our results, we expect that the importance of the SWIR domain will remain at the canopy level. Current multispectral spaceborne platforms, such as Landsat and Sentinel, do not comply with the narrowband spectral requirements we have identified. However, space-borne imaging spectroscopy data (Berger et al., 2020) are suitable for this type of N monitoring and their increasing availability is bringing new opportunities in this field. A number of satellite platforms is already operational or close to launch, e.g. PRISMA (Loizzo et al., 2019), Gaofen-5 (Liu et al., 2019), or EnMap (Guanter et al., 2015), and several candidate missions are in preparation, such as the Copernicus Hyperspectral Imaging Mission for the Environment (CHIME) (Nieke and Rast, 2018) or NASA's EMIT (Green et al., 2019) and Surface Biology and Geology (SBG) mission (Committee on the Decadal Survey for Earth Science and Applications from Space et al., 2018; Hochberg et al., 2015). Data provided by these instruments holds a strong perspective for N monitoring. Yet, preparatory studies will be necessary to analyse the potential of PROSPECT-PRO for simulating sufficiently accurate imaging spectroscopy data of canopies when being coupled with canopy RT models such as SAIL (Berger et al., 2018; Jacquemoud et al., 2009; Verhoef et al., 2007), SCOPE (van der Tol et al., 2009) or DART (Gastellu-Etchegorry et al., 2017, 2015).

7. CONCLUSIONS

This study introduces PROSPECT-PRO, a new version of the PROSPECT leaf RT model, capable of differentiating proteins from other carbon-based constituents as two independent components of LMA. The calibration of PROSPECT-PRO was based on the assumption that proteins and CBC are the two main spectrally active constituents of the leaf dry matter. We demonstrated that PROSPECT-PRO performs similarly in estimating protein and CBC content of both fresh and dry leaves, a marked improvement over previous attempts. Errors computed between measured and simulated LOP were relatively low for both types of leaves.

Our results revealed that the optimal estimation of leaf protein content at the leaf scale is obtained when using leaf optical properties from a narrow spectral domain between 2125 and 2175 nm. The estimation of protein content, assessed by NRMSE, was found to be slightly less accurate than the estimation of CBC content or total LMA. Even with these results, further investigations conducted on independent leaf-scale measurements of LOP, proteins (or nitrogen) and LMA are still needed. Canopy-scale studies are also required to test the potential of this new model for operational airborne and space-borne applications. Since current satellite multispectral instruments (e.g. MSI of Sentinel-2 or ETM on Landsat-8/9) do not possess finely enough resolved spectral characteristics suitable for estimation of vegetation protein and CBC content, spaceborne imaging spectroscopy missions may be of the critical importance for a future operational protein and related nitrogen monitoring of agricultural and natural environments.

8. ACKNOWLEDGMENTS

Authors would like to thank Philip Townsend from University of Wisconsin – Madison for his comments and advices improving scientific quality and readability of this manuscript. J.-B. Féret and F. de Boissieu acknowledge financial support from Agence Nationale de la Recherche (BioCop project—ANR-17-CE32-0001) and TOSCA program grant of the French Space Agency (CNES) (HyperTropik/HyperBIO project). Contribution of Z. Malenovský was supported by the Australian Research Council Future Fellowship 'Bridging Scales in Remote Sensing of Vegetation Stress' (FT160100477).